\documentclass[journal]{IEEEtranTIE}
\usepackage{graphicx}
\usepackage{cite}
\usepackage{picinpar}
\usepackage{amsmath}
\usepackage{url}
\usepackage{flushend}
\usepackage[latin1]{inputenc}
\usepackage{colortbl}
\usepackage{soul}
\usepackage{multirow}
\usepackage{pifont}
\usepackage{color}
\usepackage{alltt}
\usepackage[hidelinks]{hyperref}
\usepackage{enumerate}
\usepackage{siunitx}
\usepackage{breakurl}
\usepackage{epstopdf}
\usepackage{pbox}

\begin{document}
\title{A Control Oriented Fractional-Order Model of Lithium-ion Batteries Based on Caputo Definition}

\author{
Yangyang Xu, Hongyu Zhao, Chengzhong Zhang, Chenglin Liao \\
\textit{University of Chinese Academy of Sciences} \\
}

\maketitle	
\begin{abstract}
This letter proposes a fractional-order battery model based on the Caputo definition. A closed-form time-domain solution is derived, enabling a simple recursive expression for discrete-time implementation. The model requires only the current and previous time-step states in each iteration, significantly reducing memory usage compared to the conventional Gr\"{u}nwald--Letnikov (G-L) method. This recursive structure is highly compatible with filter design and online parameter identification. Experimental validation on a 40.2~Ah NCM622 cell shows that the proposed first-order model achieves voltage prediction accuracy comparable to a second-order integer-order model. The results demonstrate that the Caputo-based model offers a practical balance between accuracy and computational efficiency, making it well suited for real-time battery management systems (BMS).
\end{abstract}


% \markboth{IEEE TRANSACTIONS ON INDUSTRIAL ELECTRONICS}%
% {}

\definecolor{limegreen}{rgb}{0.2, 0.8, 0.2}
\definecolor{forestgreen}{rgb}{0.13, 0.55, 0.13}
\definecolor{greenhtml}{rgb}{0.0, 0.5, 0.0}

\section{Introduction}

\IEEEPARstart {L}{i}thium-ion  batteries have been widely applied in industrial fields due to their advantages of high energy density and high power density \cite{1}. Ensuring safe, efficient, and reliable operation requires a battery management system (BMS) capable of real--time state estimation and optimal control, for which an accurate battery model is fundamental. Currently, classical integer-order models (IOMs), such as RC circuit models, are widely adopted in industry due to their structural simplicity and low computational complexity. Dong \textit{et al.}~\cite{2} employed subspace methods to identify the model order of IOMs, and the results showed that a second-order IOM was preferred in describing battery dynamics. However, RC branches with ideal capacitors cannot fully account for the semicircle phenomenon observed in the mid-frequency region of electrochemical impedance spectroscopy (EIS)~\cite{3}. To address this limitation, fractional-order models (FOMs) have emerged in recent years. By replacing ideal capacitors with constant phase elements (CPEs), traditional IOMs are extended to FOMs, which not only improves the time domain modeling accuracy but also better simulates EIS characteristics in the frequency domain \cite{4}.

At present, most fractional-order battery models adopt the Gr\"{u}nwald--Letnikov (G-L) definition due to its inherent advantage in discretization \cite{5}. The G-L derivative is essentially a finite-difference approximation that computes a weighted sum over all historical states. While this structure facilitates numerical implementation, it inevitably requires storing $N$ past states, posing significant challenges for real-time applications. Furthermore, the G-L definition lacks explicit physical interpretability, especially in terms of initial conditions such as voltage or current, which are fundamental in battery modeling. This ambiguity in physical meaning complicates the parameter identification process, which is predominantly addressed using population-based metaheuristic algorithms such as particle swarm optimization (PSO) \cite{7} and genetic algorithm (GA) \cite{6} variants. Although these methods offer strong global search capabilities, they often suffer from long computation times and unstable identification results. In contrast, the Caputo definition is more naturally aligned with physical systems and engineering initial conditions. However, to the best of our knowledge, no existing battery model has been formulated based on the Caputo fractional derivative.

To address the aforementioned issues, this letter proposes a  fractional-order battery model based on caputo definition and a parameter identification method derived from its analytical solution. The main contributions of this work are summarized as follows:

\begin{enumerate}
    \item Starting from the Caputo definition of the fractional derivative, a rigorous closed-form derivation of the proposed model is provided. The model requires only two time-step states to be stored, significantly improving memory efficiency.
    \item Based on the analytical zero-input response of the proposed model, an efficient parameter identification method is developed. 
\end{enumerate}

\section{Preliminaries}
\subsection{Caputo Fractional Derivative}

The Caputo definition of the fractional derivative is widely adopted in engineering applications due to its compatibility with classical initial conditions \cite{8}. For a function $f(t) \in C^n[0, \infty)$, the Caputo derivative of order $\alpha \in (0,1]$ is defined as
\begin{equation}
{}^C\!D_t^\alpha f(t) = \frac{1}{\Gamma(1-\alpha)} \int_0^t \frac{f'(\tau)}{(t-\tau)^\alpha} \, d\tau
\end{equation}

Its Laplace transform is given by \cite{9}
\begin{equation}
\mathcal{L}\left\{\,{}^C\!D_t^\alpha f(t)\right\} = s^\alpha F(s) - s^{\alpha-1} f(0)
\end{equation}

where $F(s)$ is the Laplace transform of $f(t)$.

\subsection{One-Parameter Mittag--Leffler Function}

The one-parameter Mittag--Leffler function is defined by
\begin{equation}
E_{\alpha}(z) = \sum_{k=0}^{\infty} \frac{z^k}{\Gamma(\alpha k + 1)}, \quad \alpha > 0
\end{equation}

when $\alpha = 1$, this function reduces to the classical exponential function, i.e., $E_1(z) = e^z$.

\subsection{Two-Parameter Mittag--Leffler Function}

The more general two-parameter Mittag--Leffler function is given by
\begin{equation}
E_{\alpha,\beta}(z) = \sum_{k=0}^{\infty} \frac{z^k}{\Gamma(\alpha k + \beta)}, \quad \alpha > 0, \; \beta > 0
\end{equation}

when $\beta = 1$, this function reduces to the one-parameter form, i.e., $E_{\alpha,1}(z) = E_{\alpha}(z)$.

\subsection{Relation Between Mittag--Leffler Functions}

In this work, two special cases are of particular interest: $E_{\alpha}(z)$ and $E_{\alpha,\alpha+1}(z)$. By manipulating the definition of $E_{\alpha,\alpha+1}(z)$, we find
\begin{equation}
E_{\alpha,\alpha+1}(z) = \sum_{k=0}^{\infty} \frac{z^k}{\Gamma(\alpha k + \alpha + 1)} = \sum_{k=1}^{\infty} \frac{z^{k-1}}{\Gamma(\alpha k + 1)}
\end{equation}

Multiplying both sides by $z$, we obtain
\begin{equation}
z\,E_{\alpha,\alpha+1}(z) = \sum_{k=1}^{\infty} \frac{z^{k}}{\Gamma(\alpha k + 1)} = E_{\alpha}(z) - 1
\end{equation}

Thus, an important identity follows:
\begin{equation}
E_{\alpha,\alpha+1}(z) = \frac{E_{\alpha}(z) - 1}{z}.
\end{equation}

This relation plays a key role in the time-domain solution and discrete implementation developed in later sections.

\section{Modeling}
Fractional-order battery models are typically constructed by replacing conventional capacitors with constant phase elements (CPEs), which better capture the non-ideal electrochemical behavior of real cells.A CPE is characterized in the Laplace domain by the following impedance:
\begin{equation}
Z(s) = \frac{1}{C\,s^\alpha}, \quad 0 < \alpha \leq 1.
\end{equation}
\vspace{-3ex}
\begin{figure}[!t]
\centering
\includegraphics[width=\linewidth]{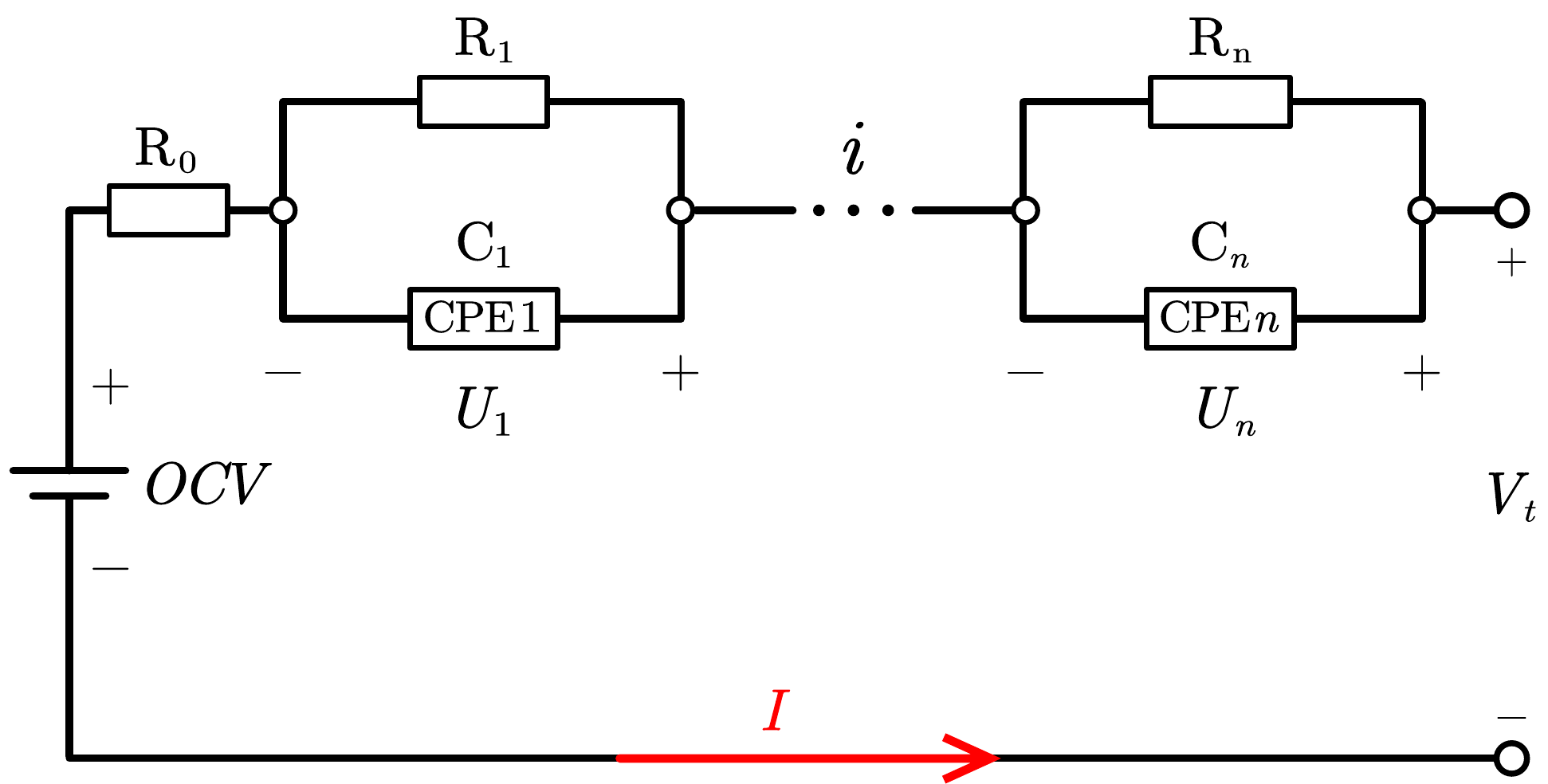}
\caption{Fractional-order nRC equivalent circuit model.}
\label{fig1}
\end{figure}

Since the RC branches are independent, we focus on the analysis of the $R_1C_1$ branch. The voltage across the CPE in this branch, denoted $U_1(t)$, satisfies the following fractional-order differential equation:
\begin{equation}
{}^C\!D_t^\alpha U_1(t) = -\frac{1}{R_1 C_1} U_1(t) + \frac{1}{C_1} I(t)
\end{equation}

Taking the Laplace transform of both sides using the Caputo derivative property in Equation (2), we obtain
\begin{equation}
s^\alpha U_1(s) - s^{\alpha-1} U_1(0) = -\frac{1}{R_1 C_1} U_1(s) + \frac{1}{C_1} I(s)
\end{equation}

which can be rearranged as
\begin{equation}
U_1(s) = \frac{s^{\alpha-1}}{s^\alpha + \frac{1}{R_1 C_1}} U_1(0) + \frac{1/C_1}{s^\alpha + \frac{1}{R_1 C_1}} I(s)
\end{equation}

The first term represents the \emph{zero-input response} due to the initial condition $U_1(0)$, while the second term is the \emph{zero-state response} caused by the input current $I(t)$.

To derive the time-domain solution, we apply the standard inverse Laplace transform identity for fractional-order systems \cite{9}:
\begin{equation}
\mathcal{L}^{-1}\left\{ \frac{s^{\alpha - \beta}}{s^\alpha + \lambda} \right\}
= t^{\beta - 1} E_{\alpha,\beta}(-\lambda t^\alpha)
\label{eq:laplace_mittag_general}
\end{equation}

For the \emph{zero-input response}, we apply Eq~\eqref{eq:laplace_mittag_general} with $\beta = 1$ and $\lambda = 1/(R_1 C_1)$, which yields
\begin{equation}
U_{1,\mathrm{zi}}(t) = U_1(0) \cdot E_\alpha\left( -\frac{t^\alpha}{R_1 C_1} \right)
\end{equation}

For the \emph{zero-state response}, we apply Eq~\eqref{eq:laplace_mittag_general}  with $\beta = \alpha$ and $\lambda = 1/(R_1 C_1)$:
\begin{equation}
\mathcal{L}^{-1}\left\{ \frac{1}{s^\alpha + \lambda} \right\}
= t^{\alpha - 1} E_{\alpha,\alpha}(-\lambda t^\alpha)
\end{equation}

Thus, the impulse response of the system is given by
\begin{equation}
h(t) = \frac{1}{C_1} \, t^{\alpha - 1} E_{\alpha,\alpha} \left( -\frac{t^\alpha}{R_1 C_1} \right)
\end{equation}

The zero-state response is then obtained by the convolution of $h(t)$ and the input current $I(t)$:
\begin{equation}
U_{1,\mathrm{zs}}(t) = \int_0^t h(t - \tau) \, I(\tau) \, d\tau,
\end{equation}

which yields the explicit expression:
\begin{equation}
U_{1,\mathrm{zs}}(t) = \frac{1}{C_1} \int_0^t (t - \tau)^{\alpha - 1}
E_{\alpha,\alpha} \left( -\frac{(t - \tau)^\alpha}{R_1 C_1} \right) I(\tau) \, d\tau
\end{equation}

In what follows, we consider a special case where the input current is constant, i.e., $I(\tau) = I_0$. Since the Mittag--Leffler function does not possess the semigroup property like the exponential function, the convolution integral cannot be simplified analytically in the usual way. To address this, we perform a change of variable by letting $z = t - \tau$, which transforms the limits to $z \in [0, t]$. The integral becomes:
\begin{equation}
U_{1,\mathrm{zs}}(t) = \frac{I_0}{C_1} \int_0^t z^{\alpha - 1}
E_{\alpha,\alpha} \left( -\frac{z^\alpha}{R_1 C_1} \right) dz
\end{equation}

This matches the structure of the following Mittag--Leffler integral identity \cite{9}:
\begin{equation}
\int_0^z t^{\beta - 1} E_{\alpha,\beta}(\lambda t^\alpha) \, dt = z^\beta E_{\alpha,\beta+1}(\lambda z^\alpha), \quad \beta > 0
\label{eq:ml_integral_identity}
\end{equation}

Applying this result with $\beta = \alpha$, $\lambda = -\frac{1}{R_1 C_1}$, and $z = t$, we obtain the closed-form expression for the zero-state response:
\begin{equation}
U_{1,\mathrm{zs}}(t) = \frac{I_0}{C_1} \, t^\alpha \, E_{\alpha,\alpha+1} \left( -\frac{t^\alpha}{R_1 C_1} \right)
\end{equation}

Applying the identity derived in Eq~(7) to convert the two-parameter Mittag--Leffler function into its single-parameter form, and letting \( z = -\frac{t^\alpha}{R_1 C_1} \), we obtain:
\begin{equation}
E_{\alpha,\alpha+1}\left( -\frac{t^\alpha}{R_1 C_1} \right)
= \frac{R_1 C_1}{t^\alpha} \left[ 1 - E_\alpha\left( -\frac{t^\alpha}{R_1 C_1} \right) \right]
\end{equation}

Substituting this into the Eq~(20) gives:
\begin{equation}
U_{1,\mathrm{zs}}(t)
= I_0 R_1 \left[ 1 - E_\alpha\left( -\frac{t^\alpha}{R_1 C_1} \right) \right]
\end{equation}

At this point, we have obtained the complete time-domain response of the fractional-order system under a constant current input:
\begin{equation}
U_1(t) = U_1(0) \cdot E_\alpha\left( -\frac{t^\alpha}{R_1 C_1} \right)
+ I_0 R_1 \left[ 1 - E_\alpha\left( -\frac{t^\alpha}{R_1 C_1} \right) \right]
\end{equation}

Notably, the response is fully expressed in terms of the one-parameter Mittag--Leffler function, enabling efficient evaluation without numerical convolution.

When $\alpha = 1$, the response degenerates into the classical integer-order solution:
\begin{equation}
U_1(t) = U_1(0) \cdot e^{-\frac{t}{R_1 C_1}} + I_0 R_1 \left( 1 - e^{-\frac{t}{R_1 C_1}} \right)
\end{equation}

Interestingly, this is exactly the analytical solution of the standard integer-order 2RC model. This observation reinforces the interpretation of fractional calculus as a generalization of classical calculus, rather than an entirely separate formulation.

To implement the model numerically, we adopt a zero-order hold (ZOH) assumption where the input current remains constant in each sampling interval. Let $T$ be the sampling period and define $U_{1,k} = U_1(kT)$, $I_k = I(kT)$. Using the closed-form response under constant current, the voltage recursion at sampling points is:
\begin{equation}
U_{1,k+1} = U_{1,k} \cdot E_\alpha\left( -\frac{T^\alpha}{R_1 C_1} \right)
+ I_k R_1  \left[ 1 - E_\alpha\left( -\frac{T^\alpha}{R_1 C_1} \right) \right]
\end{equation}

Compared to the Gr\"{u}nwald--Letnikov (G-L) discretization, the proposed recursion requires storing only two time steps, avoiding the need to retain the $N$-step history. The main computational cost lies in evaluating the Mittag--Leffler function. To accelerate this process, we adopt a dynamic truncation strategy based on the following series expansion:
\begin{equation}
E_\alpha(z) \approx \sum_{k=0}^N \frac{z^k}{\Gamma(k\alpha + 1)}, \quad 0 < \alpha \le 1
\end{equation}

Since the Gamma function grows rapidly with $k$, the contribution of higher-order terms diminishes quickly. In practice, we terminate the summation once the absolute value of the $k$-th term satisfies:
\begin{equation}
|\text{Term}_k| < \epsilon
\end{equation}

In practical applications, the tolerance $\epsilon$ can be adjusted to meet varying accuracy requirements under different scenarios.In this work, the tolerance is set to $\epsilon = 10^{-6}$.

By combining the ampere-hour counting method and Kirchhoff's voltage law, we derive the following discrete-time state and observation equations for the system.

\textbf{State equation:}
\begin{equation}
\begin{bmatrix}
SOC_{k+1} \\
U_{1,k+1} \\
\vdots \\
U_{n,k+1}
\end{bmatrix}
=
\begin{bmatrix}
1 & 0 & \cdots & 0 \\
0 & a_1 & \cdots & 0 \\
\vdots & \vdots & \ddots & \vdots \\
0 & 0 & \cdots & a_n
\end{bmatrix}
\begin{bmatrix}
SOC_k \\
U_{1,k} \\
\vdots \\
U_{n,k}
\end{bmatrix}
+
\begin{bmatrix}
b_0 \\
b_1 \\
\vdots \\
b_n
\end{bmatrix} I_k
\end{equation}

where $a_i = E_{\alpha_i}\left( -\dfrac{T^{\alpha_i}}{R_i C_i} \right)$, $b_i = R_i \left( 1 - a_i \right)$ and $b_0 = \dfrac{T}{Q_n}$

\textbf{Observation equation:}
\vspace{-2ex}
\begin{equation}
V_{t,k+1} = OCV(SOC_{k+1}) + \sum_{i=1}^{n} U_{i,k+1} + R_0 I_{k+1}
\end{equation}

\section{Experiment}
\subsection{Experimental setup}
In this letter, a commercial automotive lithium-ion battery (NCM622, 40.2~Ah) is selected for testing. The experimental setup is shown in Fig.~2. To identify model parameters, a series of Hybrid Pulse Power Characterization (HPPC) tests were conducted at 5\% SOC intervals. In addition, the standard Federal Urban Driving Schedule (FUDS) profile was employed to validate the effectiveness of the proposed model under realistic operating conditions. The open-circuit voltage (OCV) curve was obtained by averaging the voltage data from 0.05C charging and 0.05C discharging steps.
\begin{figure}[!h]
\centering
\includegraphics[width=\linewidth]{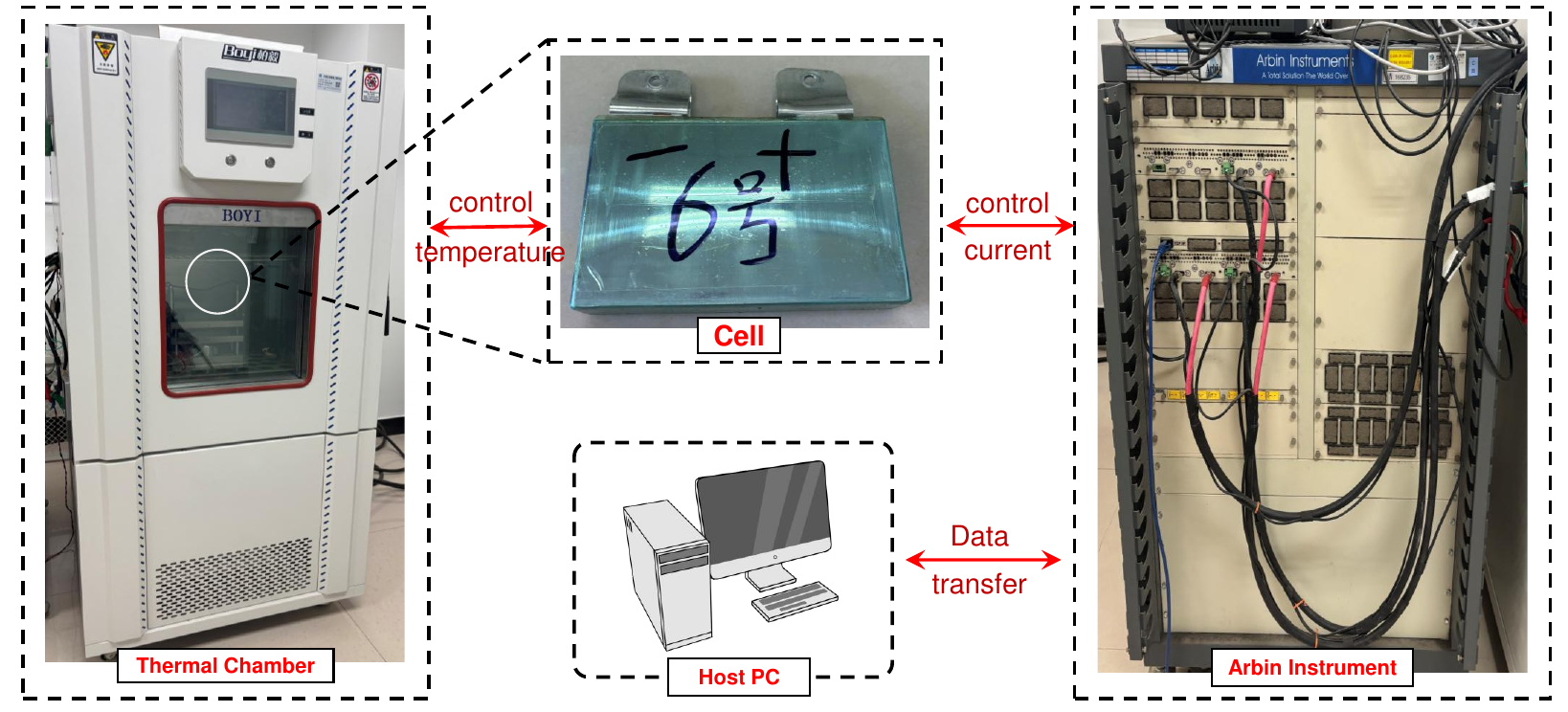}
\caption{ The experimental platform}
\label{fig2}
\end{figure}

\vspace{-2ex}
\subsection{Parmeters Identification}
\begin{figure}[!h]
\centering
\includegraphics[width=\linewidth]{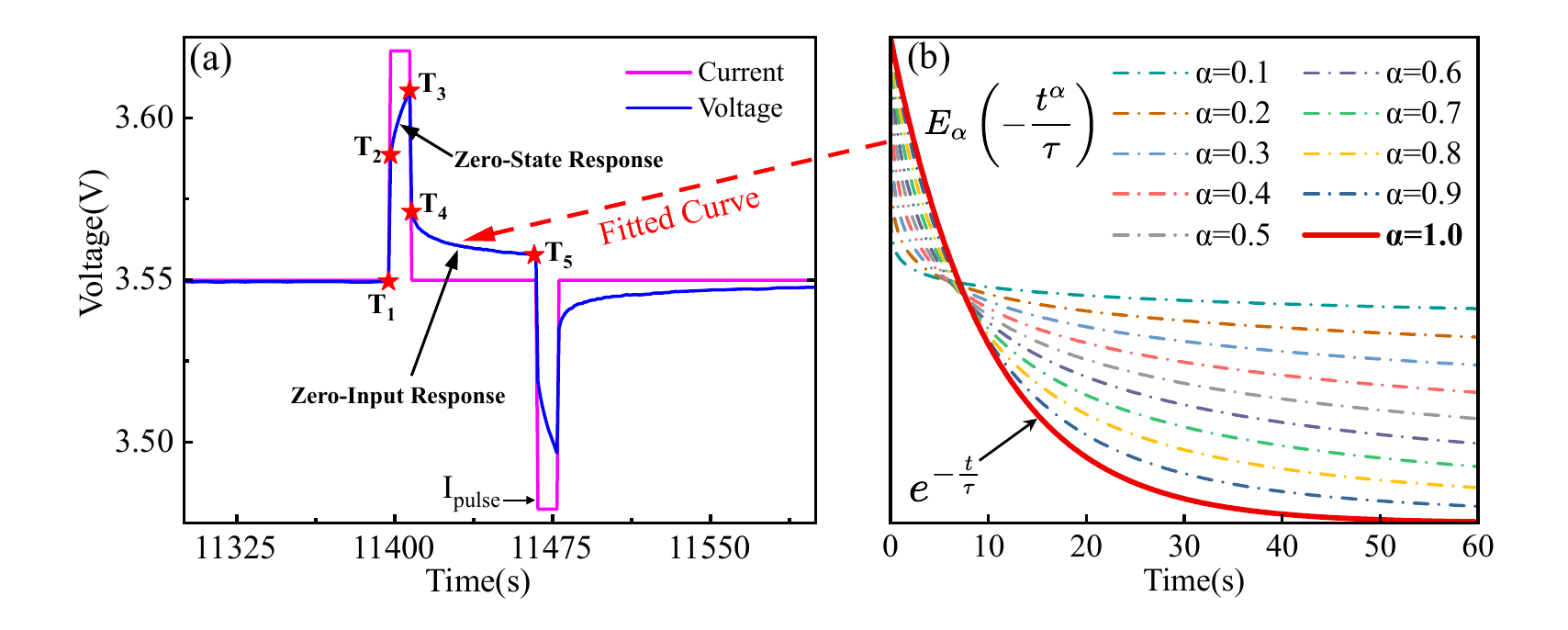}
\caption{(a) Single HPPC test segment illustrating zero-state responses. (b) Variation of Mittag--Leffler function $E_{\alpha}\left( -\frac{t^{\alpha}}{\tau} \right)$ with different $\alpha$.}
\label{fig3}
\end{figure}
Since the analytical solution of the zero-state response has been derived in Section III, it is natural to consider using HPPC data for parameter identification.

The ohmic resistance $R_0$ is first calculated based on the voltage transient in Fig.~3 (a), as given by:
\begin{equation}
R_0 = \frac{U_{T_2} - U_{T_1} + U_{T_3} - U_{T_4}}{2 I_{\mathrm{pulse}}}
\end{equation}

Next, the fractional-order parameters are identified by fitting the zero-state voltage response observed during the relaxation period ($T_4$ to $T_5$) using the following analytical expression:
\begin{equation}
V_t(t) = {OCV}(SOC_j) + I_{\mathrm{pulse}} \sum_{i=1}^{n} R_i \, E_{\alpha_i} \left( -\frac{t^{\alpha_i}}{\tau_i} \right)
\end{equation}

The model parameters are estimated by minimizing the mean squared error between the measured and predicted voltages. In this letter, \textit{Trust-Region Reflective Method} is adopted to handle the nonlinear least-squares problem with bound constraints effectively.

\subsection{Results and Discussion}
As illustrated in Fig.~3 (b), the Mittag--Leffler function exhibits a broader range of behaviors than the conventional exponential function, thereby providing enhanced modeling flexibility for complex electrochemical phenomena.
\begin{figure}[!h]
\centering
\includegraphics[width=\linewidth]{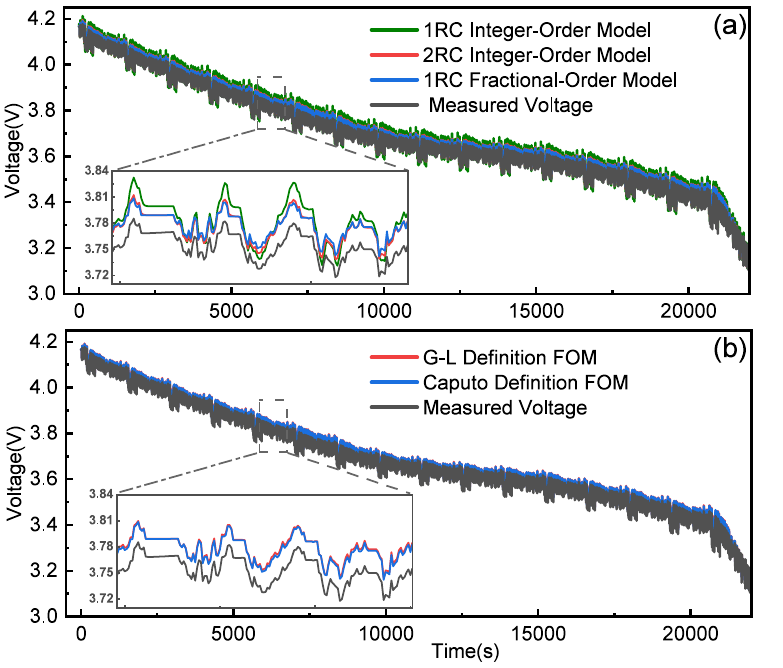}
\caption{Voltage estimation under the FUDS profile. (a) Comparison between the 1RC fractional-order model and integer-order models. (b) Comparison between Caputo-based and G-L-based fractional-order models.}
\label{fig4}
\end{figure}

As shown in Fig.~4 (a), the 1RC FOM achieves voltage estimation performance comparable to the 2RC IOM. This is consistent with the earlier analysis that a single fractional-order RC branch can accurately capture the zero-input response during the relaxation period ($T_4$ to $T_5$), whereas two integer-order RC branches are needed for a similar fit. Unlike the classical integer-order models that are limited to exponential responses, fractional-order models extend the dynamic representation capability by introducing the fractional order $\alpha$ as an additional degree of freedom. 

As shown in Fig.~4 (b), the proposed fractional-order model based on the Caputo definition achieves voltage prediction accuracy nearly identical to that of the Gr\"{u}nwald--Letnikov (G-L) based model. This is because, mathematically, the two definitions are equivalent. However, the Caputo-based model requires storing only two previous time steps in each iteration, whereas the G-L-based model must access the past $N$ steps according to the finite memory principle. In this work, $N$ is set to 64. This substantial reduction in computational burden underscores the efficiency advantage of the Caputo formulation for real-time applications.

\section{Conclusion}
This letter proposes a control-oriented fractional-order model based on the Caputo definition, along with a rigorous closed-form derivation of its time-domain response and discrete-time implementation. Compared to the widely used G-L discretization, the Caputo formulation retains only two historical states per iteration, offering a fixed low-memory requirement well suited for real-time applications. Furthermore, its recursive structure is inherently compatible with filter design and online parameter identification. In summary, the Caputo-based fractional-order model achieves a favorable balance between accuracy and computational efficiency, supporting its practical deployment in onboard battery management systems. 

% References

\bibliographystyle{IEEEtranTIE}

\ %IEEEabrv instead of IEEEfull

\end{document}